\documentclass[preprint,pre,a4paper,superscriptaddress,showpacs,nofootinbib]{revtex4}
 
\usepackage{amsmath}
\usepackage{amsfonts}
\usepackage{amssymb}
\usepackage{bm}
\usepackage{latexsym}
\usepackage{isolatin1}

\usepackage{graphicx}

\usepackage{float}

\newcommand{\la}{\langle}
\newcommand{\ra}{\rangle}

\newcommand{\usvec}{\mbox{$\underline{u}_s$}}
\newcommand{\fsvec}{\mbox{$\underline{f}_s$}}

\newcommand{\FNvec}{\mbox{$\underline{F}$}}
\newcommand{\UNvec}{\mbox{$\underline{U}$}}
\newcommand{\sigmaxx}{\mbox{$\sigma_{xx}$}}
\newcommand{\sigmayy}{\mbox{$\sigma_{yy}$}}
\newcommand{\sigmaxy}{\mbox{$\sigma_{xy}$}}

\newcommand{\Mtensor}{\mbox{\underline{\underline{M}}}}

\newcommand{\rsfig}[1]{\begin{center}
                       \includegraphics*[width=0.8\textwidth]{{#1}}
                       \end{center}
                       }

\begin{document}
\sloppy

\title{ Continuum limit of amorphous elastic bodies (II):\\ 
Linear response to a point source force}

\author{F. Leonforte}
\affiliation{Laboratoire de Physique de la Mati\`ere Condens\'ee et des Nanostructures,
Universit\'e Claude Bernard (Lyon I) \& CNRS, 43 Bvd. du 11 Nov. 1918, 
69622 Villeurbanne Cedex, France}

\author{A.~Tanguy}
\email[Email: ]{atanguy@lpmcn.univ-lyon1.fr}
\affiliation{Laboratoire de Physique de la Mati\`ere Condens\'ee et des Nanostructures,
Universit\'e Claude Bernard (Lyon I) \& CNRS, 43 Bvd. du 11 Nov. 1918, 
69622 Villeurbanne Cedex, France}

\author{J.P.~Wittmer}
\email[Email: ]{jwittmer@ics.u-strasbg.fr}
\affiliation{Institut Charles Sadron, 6, Rue Boussingault, 67083 Strasbourg, France}

\author{J.-L. Barrat}
\affiliation{Laboratoire de Physique de la Mati\`ere Condens\'ee et des Nanostructures,
Universit\'e Claude Bernard (Lyon I) \& CNRS, 43 Bvd. du 11 Nov. 1918, 
69622 Villeurbanne Cedex, France}

\begin{abstract}
The linear response of two-dimensional amorphous elastic bodies to an external delta force is determined in analogy with recent experiments on granular aggregates. For the generated forces, stress and displacement fields, we find strong relative fluctuations of order one close to the source, which, however, average out readily to the classical predictions of isotropic continuum elasticity. 
The stress fluctuations decay (essentially) exponentially with distance from the source. Only beyond a surprisingly large distance, $b \approx 30$ interatomic distances, self-averaging dominates, and the quenched disorder becomes irrelevant for the response of an individual configuration. 
We argue that this self-averaging length $b$ sets also the lower wavelength bound for the applicability of classical eigenfrequency calculations.
Particular attention is paid to the displacements of the source, allowing a direct measurement of the local rigidity. 
The algebraic correlations of these displacements demonstrate the existence of domains of slightly different rigidity without, however, revealing a characteristic length scale, at least not for the system sizes we are able to probe.

\end{abstract}


\maketitle

\section{Introduction.}
\label{sec:introduction}

The recent years have seen a tremendous effort to determine the response of granular matter subject to point (delta) sources as indicated in Fig.~\ref{figsketch}(a). These theoretical~\cite{forcechains,chaos,claudinaniso,gay,GG01}, experimental~\cite{liunagel,claudinelastic,clement,Vanel,Kolb} and computational~\cite{radjai} studies have been motivated by the desire to understand the static properties of, say, a humble sandpile --- to quote an important paradigmatic example~\cite{chaos}. 
It has been argued that these aggregates formed under gravity as external driving force --- alongside with other special ``solids'' such as jammed colloids, emulsions or foams --- may not necessarily be described as classical elastic or elastoplastic continuum bodies~\cite{chaos,GG01,radjai}. 
Hence, the interest to determine experimentally and by computer simulation the linear and quasistatic response to a localized incremental force, in order to distinguish between the different models proposed.
In a nutshell, stress distributions below the source rather close to classical elasticity predictions have been found for standard sand, although minor differences seem to appear in the distribution tails \cite{clement}. 
This has prompted the more recent focus on the fact that these systems are typically composed of a small number of constituents \cite{GG01}, and on the paramount role of the quenched disorder \cite{claudinaniso}. 

In this paper, the point source response problem is carried over to a definitely much simpler disordered model system, the two-dimensional amorphous solid formed by quenching a Lennard-Jones fluid. 
It is well known for amorphous materials such as metallic, organic or mineral glasses, that their mechanical properties are quite different from those of the corresponding crystals at the same density~\cite{amorphes,masumoto}. They are characterized by a large decrease in both the apparent shear and Young's moduli, and a large increase of the yield stress associated with a localization of the plastic deformation~\cite{amorphes,masumoto}. These properties have been interpreted in terms of local rearrangements~\cite{Weaire,cavaille,Das,Parisi} due to the heterogeneity of the microscopic structure. But these rearrangements have never been identified clearly. Particularly, like in granular materials, the role of the quenched stresses is actually a matter of debate~\cite{Alexander,Kusta}, as well as the role of local heterogeneities in the elastic constants of the materials. 
One way to answer those questions experimentally is to perform nanoscale indentation~\cite{Loubet}, that is to study the response to a point force.

As schematically illustrated in Fig.~\ref{figsketch}~(b) we study stress and displacement fields generated by an external force acting on the Lennard-Jones beads contained within a small disk. Snapshots of the incremental stresses and displacement fields presented in Figs.~\ref{figsnapforce}, \ref{figsnapdeltau} and \ref{figDepStress} show a rather noisy response.  
Since these systems behave clearly as classical elastic bodies --- provided sufficiently large wave lengths and small forces are probed --- they provide important reference systems, the results from granular matter may be compared with.
The linear response to point source is equivalent to the (noisy and position dependent) Green's function.
To our knowledge this study presents the first systematic computational study of this function, extending some aspects discussed only recently \cite{GG01}. 
Specifically, we investigate how this noise affects the averages compared to classical continuum theory, and how the distributions get narrower with increasing distance from the source, due to self-averaging. 
The spatial correlations of the responses of close sources are studied in order to verify whether domains of different rigidity exist as has been argued recently~\cite{Alexander,Kusta,Duval}. 

This work is in fact the natural sequel of our study \cite{papAmorph0,papAmorph1} where the applicability of classical elastic continuum theory on small length scales has been tested by comparing with theory the low end of the eigenfrequency spectrum obtained by diagonalization of the dynamical matrix. We found that only for system sizes and wave lengths larger than a characteristic wave length, $\xi \approx 30$ interatomic distances, the eigenfrequencies show the degeneracy predicted for a classical isotropic and homogeneous body. 
This surprising large lower limit for classical continuum theory is also seen in the characterization of the {\em non-affine} field generated by macroscopic deformations (shear or elongation). Only after coarse-graining over distances of order $\xi$ does the non-affine response become negligible. This does in turn explain why modes associated with smaller wave lengths do not behave as predicted from an approach formulated in terms of affine displacement fields. 

In this paper, we describe first briefly some technical points related to the initial samples, the computational methods and measurements. In the subsequent Sections~\ref{sec:resBY1}, \ref{sec:resBY2} and \ref{sec:resBY3}, we present our numerical results for stress and displacement fields, and their distributions. We have regrouped our results following the three different boundary conditions investigated.
In the first section, we demonstrate that the self-averaging is characterized by a length scale similar to the critical wave length $\xi$ from our previous study. In the latter two sections we analyze the source displacements and their correlations. Our results are summarized in Sec.~\ref{sec:conclusion}. The analytical predictions from classical elasticity theory are outlined in the appendix.

\section{Computational technicalities.}
\label{sec:technicalities}

The initial configurations and their preparation have been to some extent described in \cite{papAmorph1}. Of relevance here is a large ensemble of 16 independent configurations containing each 10,000 Lennard-Jones particles quenched from $T=1$ down to zero temperature following a fixed protocol using standard molecular dynamics, steepest descent and conjugate gradient methods \cite{Thijssen}. Note that, while the particle mass $m$ is strictly monodisperse, we use sufficiently polydisperse particle diameters (uniformly distributed between $0.8$ and $1.2$) to prevent crystalline order.
The linear size of the periodic boxes is $L=104$, the corresponding volume fraction 0.925. The mean pressure $P=0.25$ was chosen to be close to zero. 
The two Lam\'e coefficients \cite{landau,Salencon,papAmorph1}, $\lambda \approx 39.5$ and $\mu \approx 11.7$, have been measured directly using Hooke's law by applying macroscopic elongation and (pure) shear to the simulation box. We recall that the associated Poisson ratio $\nu = \lambda/(\lambda+2\mu) \approx 2/3$ is larger than $1/2$ which is permissible in a 2D solid. Here as everywhere below we have naturally given the numerical values in Lennard-Jones units.

It has been carefully checked that the initial configurations and their monomers are indeed at mechanical equilibrium, i.e. are sitting in (local) minima of the energy landscape. The {\em linear} response to a small external force or imposed displacement can, hence, be described by means of the $(2N)\times(2N)$ dynamical matrix \Mtensor \ whose elements depend on the frozen tensions (``quenched stresses") and stiffnesses of the links between interacting beads \cite{papAmorph1}. In principle, it is straightforward to solve numerically the linear equations $\Mtensor \cdot \UNvec = \FNvec$.
Here, \FNvec \ and \UNvec \ are the $2N$-dimensional force and displacement fields respectively containing the imposed external body forces and displacements. Since we are considering very large systems and standard linear equation solver being of order $N^3$ we have mainly used (Sec.~\ref{sec:resBY1} and \ref{sec:resBY2}) direct steepest descent and/or conjugate gradient methods which are in this case (where the neighbor contact lists remains constant) of order $N$.
The advantage of the direct methods is also that they allow to probe the non-linear response regime. We have checked that both methods yield the same results for sufficiently small external forces. For comparison, we present in Sec.~\ref{sec:resBY3} results obtained directly from the dynamical matrix. 
    
In all cases, as shown in Fig.~\ref{figsketch}(b), we apply a localized external force of $\fsvec/n_0$ to all the $n_0$ beads contained in small source disks of fixed diameter $D$. The center of the disk refers naturally to the origin of our coordinate system $(x,y)$. 
The special limit with sources containing only one bead ($n_0=1,D \rightarrow 1$) will be used in Sec.~\ref{sec:resBY3}.
Obviously, the response becomes locally less noisy with increasing source size.
As we are interested in disorder on distances larger than the typical particle distance we have also distributed the source over more than one bead.
Most of the results reported in Sections~\ref{sec:resBY1} and \ref{sec:resBY2} are for $D=4$ corresponding to $\la n_0 \ra \approx 12$ beads. 
All the source forces considered here point vertically downwards. 
It turns out that an applied force of order one per bead is sufficiently small to ensure linear elastic response for the direct minimization methods. (See Fig.~\ref{figstressD} below.) 
The averages are taken over different disk positions in the same configuration, and also over the configuration ensemble. 

For mechanical stability, we have either imposed a compensation force of $-\fsvec/N$ on all beads or fixed the positions of certain beads, as shown in Fig.~\ref{figsketch}(b).
The first method has the advantage of being free of any fixed boundary layer making it possible to use the full initial periodic box. Care has to be taken however, in this case, for numerical reasons, because small drifts of the system cannot be completely avoided. The displacement fields must thus be considered in the center of mass frame. Sec.~\ref{sec:resBY2} presents results averaged over nearly 4000 linear responses obtained with this boundary condition. 

Most of the work presented in this paper (Sec.~\ref{sec:resBY1},~\ref{sec:resBY3}) uses instead fixed beads to compensate the source force. Either we fix all beads in a horizontal layer with $|y| > h$ and $h \le L/2$ (Sec.~\ref{sec:resBY1}), or all beads which are beyond a given number of topological layers around 
the source particles (Sec.~\ref{sec:resBY3}). 

It is well known for elastic bodies in two dimensions that stresses and strains far from both source and boundary decrease inversely with the source distance $r$, i.e. the displacement field varies logarithmically. 
Obviously, the response depends generally on the imposed boundary conditions. 
Details of the rigorous analytical treatment, exemplified for the boundary conditions studied in the next section, are outlined in the appendix.

\section{Results for fixed top and bottom layers.}
\label{sec:resBY1}

\begin{figure}[t]
\rsfig{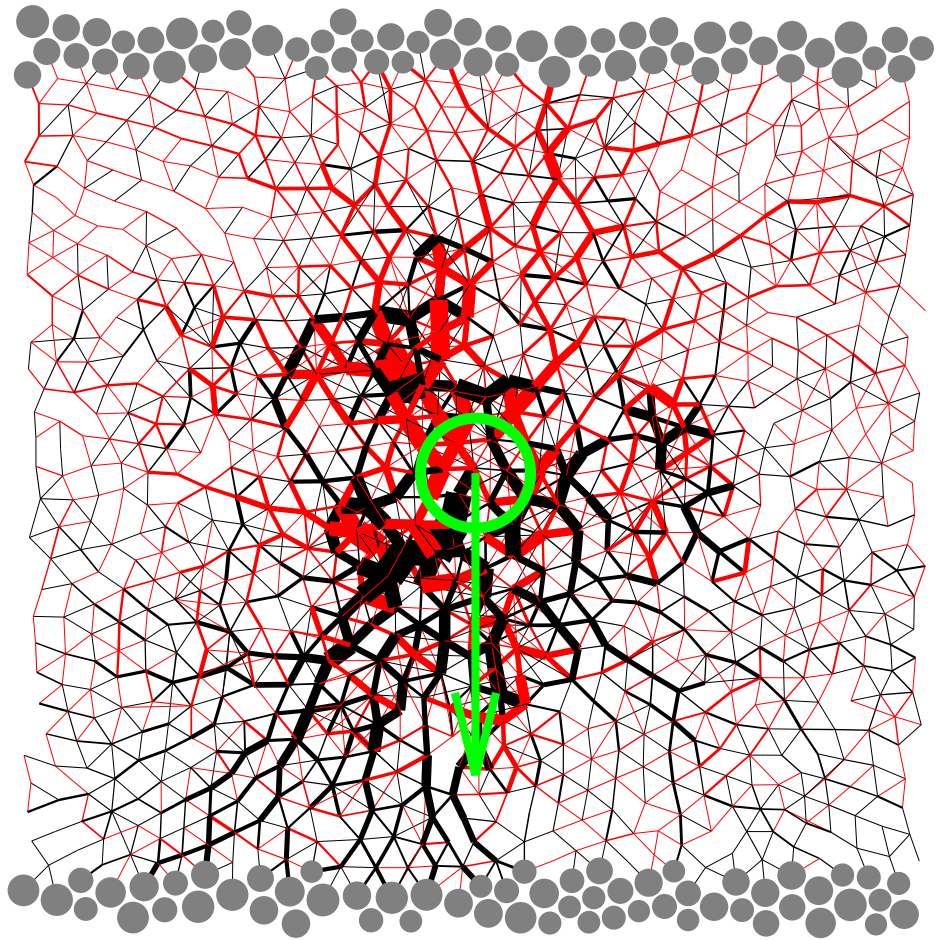}
\vspace*{0.8cm}
\caption[]{(Color online)
Snapshot of the incremental forces in one small periodic box containing $N=1000$ particles generated by the source applied on all the beads within the disk indicated. We have chosen here a disk diameter of 4 particle sizes. The line width between interacting beads is proportional to the incremental forces. (Only forces larger than 0.02 have been drawn for clarity.) Black (gray) lines correspond to incremental compressive (tensile) stresses. Also indicated on top and bottom are the beads fixed to balance the source. 
The snapshot shows that the forces generated by one additional source are strongly heterogeneous and resemble qualitatively the ``force chains" known from granular matter \cite{chaos}.
\label{figsnapforce}}
\end{figure}

\begin{figure}[t]
\rsfig{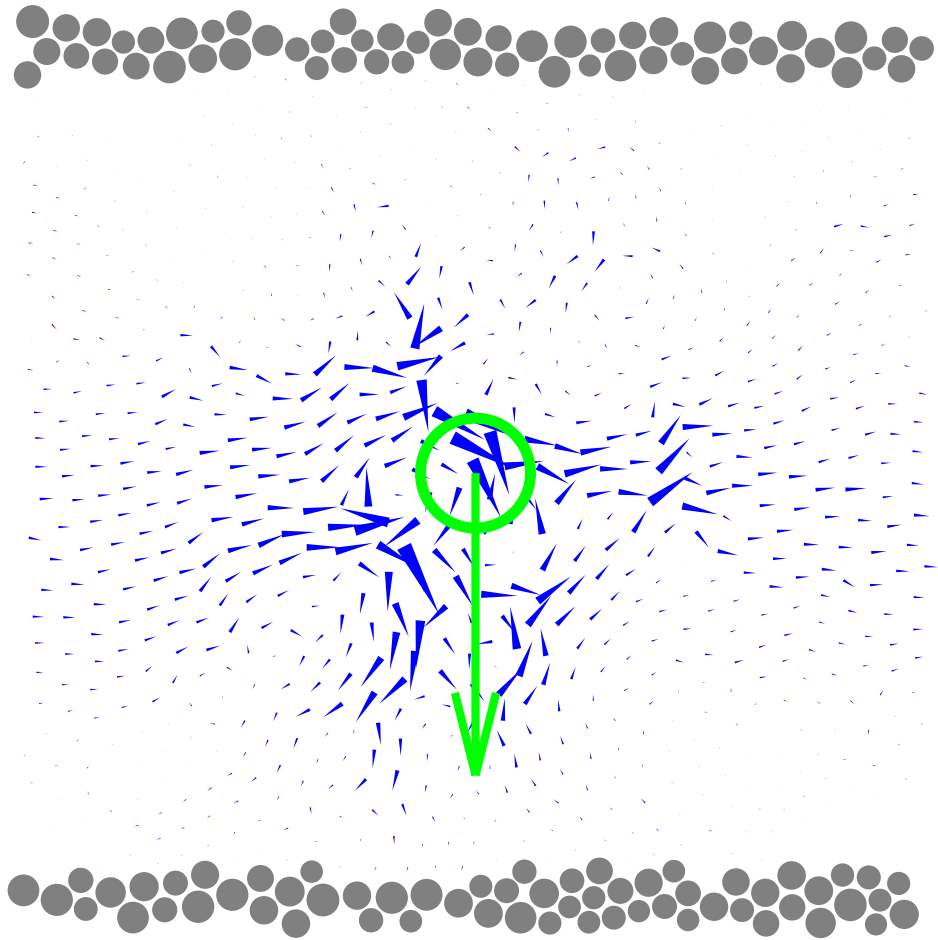}
\vspace*{0.8cm}
\caption[]{(Color online)
Snapshot of the (reduced) displacement field $\delta u = u-u_{cet}$ generated in the same configuration as in Fig.~\protect\ref{figsnapforce}. We have substracted from the total displacement field $u$, the displacement field $u_{cet}$ obtained analytically from classical continuum elasticity theory (CET). The difference from continuum theory is quite marked for the displacement of beads on the ``force chains" of Fig.~\protect\ref{figsnapforce}. On larger distances rotatory structures become visible --- quite similar to the ones obtained from the non-affine part of the displacement fields under macroscopic strain \cite{papAmorph1}. 
\label{figsnapdeltau}}
\end{figure}

The two snapshots of the forces and displacement fields depicted in Fig.~\ref{figsnapforce} and \ref{figsnapdeltau} show the response fields obtained in a small system of linear size $L=32.8$ containing only $N=1000$ beads, but at the same volume fraction and pressure as the larger samples studied quantitatively below.
The strength of the forces between beads are represented in Fig.~\ref{figsnapforce} by the width of the lines repulsive (tensile) forces being black (gray). Only the incremental forces $\delta f$ due to the source are given, i.e. the rather strong quenched or residual forces of the amorphous body have been subtracted. The force chains visible resemble strongly the ones known from granular matter ~\cite{forcechains,GG01,radjai}, although our system is certainly a classical isotropic elastic body at large distances~\cite{papAmorph1}.

\begin{figure}[t]
\rsfig{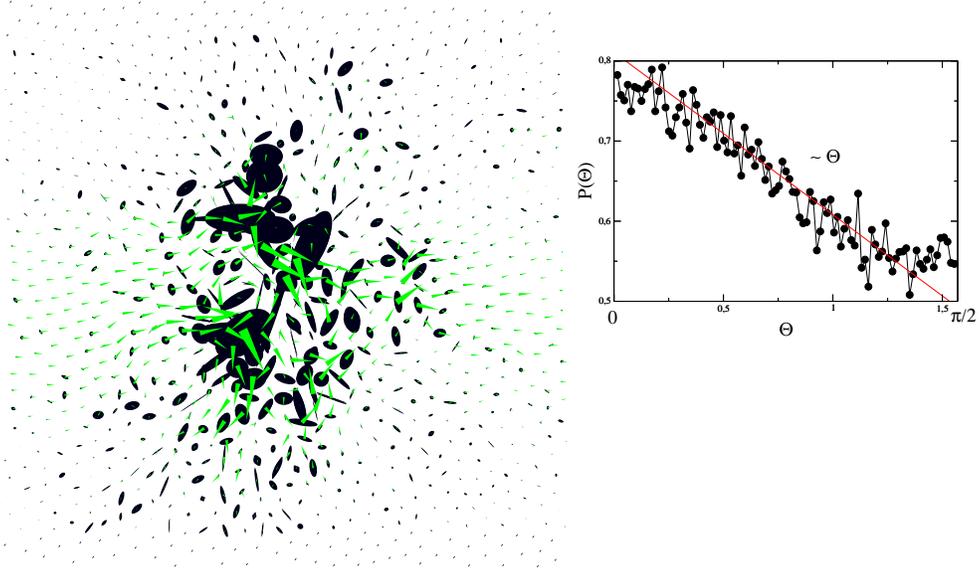}
\vspace*{0.8cm}
\caption[]{(Color online)
LEFT: Snapshot of the (reduced) displacement field $\delta u = u-u_{cet}$ superimposed with the noise component of the incremental stresses (that is $\sigma - \sigma_{quenched} - \sigma_{cet}$). The chosen configuration is the same as in Fig.~\protect\ref{figsnapforce} and Fig.~\protect\ref{figsnapdeltau}. Stresses are represented by ellipses whose large principal axis is proportional to the largest eigenvalue of the local (incremental) stress tensor. The small axis is proportional to the smallest eigenvalue of this stress tensor. The directions of the axes of the ellipses give the main directions of stress. The arrows represent the displacement field, as in Fig.~\ref{figsnapdeltau}.
  
RIGHT: Histogram of the angles $\theta$ between the local (reduced) displacement, and the main direction of the (incremental) stress tensor. The histogram is peaked around zero, with a broad distribution $\propto \theta$. It has been obtained from 10 configurations of $N = 10 000$ particles.
\label{figDepStress}}
\end{figure}

The displacement field $\delta u = u - u_{cet}$ indicated by the arrows in Fig.~\ref{figsnapdeltau} has been obtained by substracting from the total displacement field $u$, the displacement field $u_{cet}$ calculated for standard continuum elasticity theory (CET) \cite{landau,Salencon} following the prescription indicated in the appendix.  In other words, $\delta u$ depicts the noisy response due to the quenched disorder. 
In order to do a comparison of both snapshots, we show in Fig.~\ref{figDepStress} the residual displacement field $\delta u$, and the noise component of the local incremental stress on each particle. In order to obtain the noise component, we have substracted the stress calculated with standard continuum elasticity theory, and the quenched stresses, from the total stress in the presence of a source. The total stress has been calculated here on each particle, using the standard Kirkwood definition \cite{GG01,Alexander}. The noisy part of the incremental stress is then represented by an ellipse centered on the particle, whose {\it large} principal axis is proportional to the largest eigenvalue, and whose {\it small} principal axis is proportional to the smallest eigenvalue of the residual stress tensor. The directions of these axes give thus the main directions of the incremental stresses. The snapshot of Fig.~\ref{figDepStress} shows clearly that $\delta u$ is corrrelated to the local incremental stress. To get more quantitative results, we have drawn in the inset of Fig.~\ref{figDepStress} the distribution of the angles $\theta$ between the residual displacements $\delta u$, and the main direction of the incremental stresses (the direction associated to the largest eigenvalue). We show a peak for zero angle, with a broad distribution (linear with $\theta$). The residual displacement field thus reveals a clear tendency to align with the main direction of the incremental stresses.
On larger distances, however, we see a vortex like structure for $\delta u$ similar to the structure revealed by the non-affine displacement field under macroscopic strain found in \cite{papAmorph1}. The reason for this can be easily understood for the latter case where the pressure must become macroscopically constant, and with it the particle density as well. 
This generates the ``backflow" of the non-affine displacement, just like in a uncompressible fluid. We recall that the continuum displacement field $u_{cet}$ for the point source problem flows also back, but on a distance $L/2$ given by the system size. The size of the vortices measured in $\delta u$, however, does not depend on the system size. 
Note that, from one configuration to the next, the vortices are not located at the same place, and that they disappear after averaging the displacement field over many configurations.
Moreover, these vortices are not due to the natural discretization of our system: they would disappear if the spatial distribution of atoms were ordered, as can be infered from the direct computation of the Green function (see for example Ref. \cite{Green3D}).
Note finally that the 3D case is now under study. We would not be surprised if the size of the structure involved in the local rearrangements of the atoms were smaller in the 3D case, due to the minor effect of disorder and the smaller range of elasticity \cite{Ziman}. However, systems with a {\it very} large number of atoms have to be studied in this case \cite{papAmorph1} to fit with the continuum limit.

\begin{figure}[t]
\begin{center}\includegraphics*[width=0.9\textwidth]{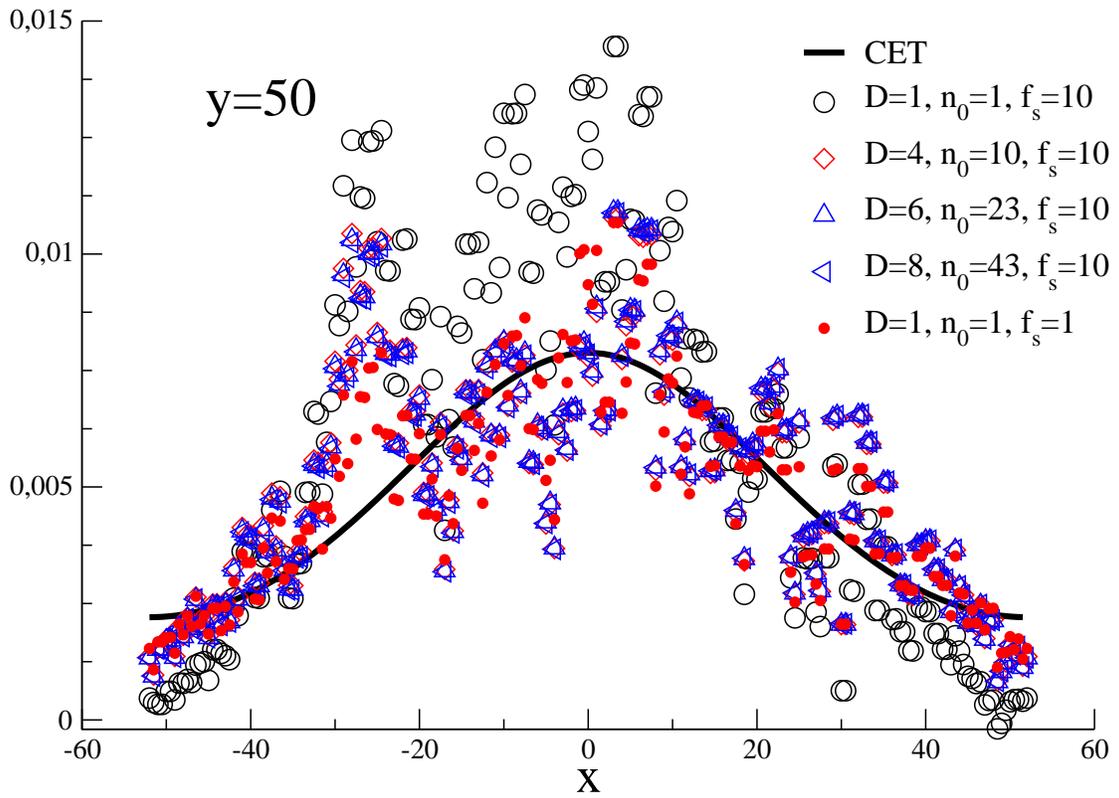}\end{center}
\vspace*{0.8cm}
\caption[]{(Color online)
Unaveraged vertical stress \sigmayy \ close to the bottom plate ($y=50$) caused by sources of various disk diameters $D$ (as indicated in the figure) at the same position of one configuration of linear size $L=104$. The total vertical stress has been used as a normalization.
The open symbols correspond to a total applied force $f_s =10$, the filled circles are for a source with $D=n_0=1$ and $f_s/n_0=1$. The bold line shows the theoretical prediction. It corresponds also to the statistical average (see Fig.~\ref{figstressmean}). The linear responses for $D=4,6$ and $8$ are perfectly identical. This only applies as long as the force per bead remains sufficiently small. 
\label{figstressD}}
\end{figure}

Fig.~\ref{figstressD} shows the vertical normal stress \sigmayy \ generated by {\em one} source of diameter $D$, at a distance $y=50$ below the source, i.e. just above the fixed beads of the bottom layer. As in the snapshot Fig.~\ref{figsnapforce}, only the incremental stresses due to the source are shown here. 
To make comparison between the sources of different strengths, the total vertical stress has been normalized to $1$.

The  stress tensor has been measured, as everywhere in the following, by means of the virial definition \cite{papAmorph1} averaged over all beads contained in small rectangular volume elements of width $5$ and height $3$ centered at $(x,y)$. Adopting in this work the sign convention usual in granular matter, compressive stresses are taken as positive, i.e. have the same sign as the pressure.
The size and the aspect ratio of the volume elements were chosen for convenience. A typical volume element contains 14 beads, and averages over about 100 interactions which takes out some of the noise. On the other hand, it remains small enough to achieve a good spatial resolution.
Note that a given interaction may contribute to two neighboring volume elements. Data points corresponding to two such elements are therefore statistically correlated, and the curves appear slightly smoother as they would otherwise. 

\begin{figure}[t]
\begin{center}\includegraphics*[width=0.9\textwidth]{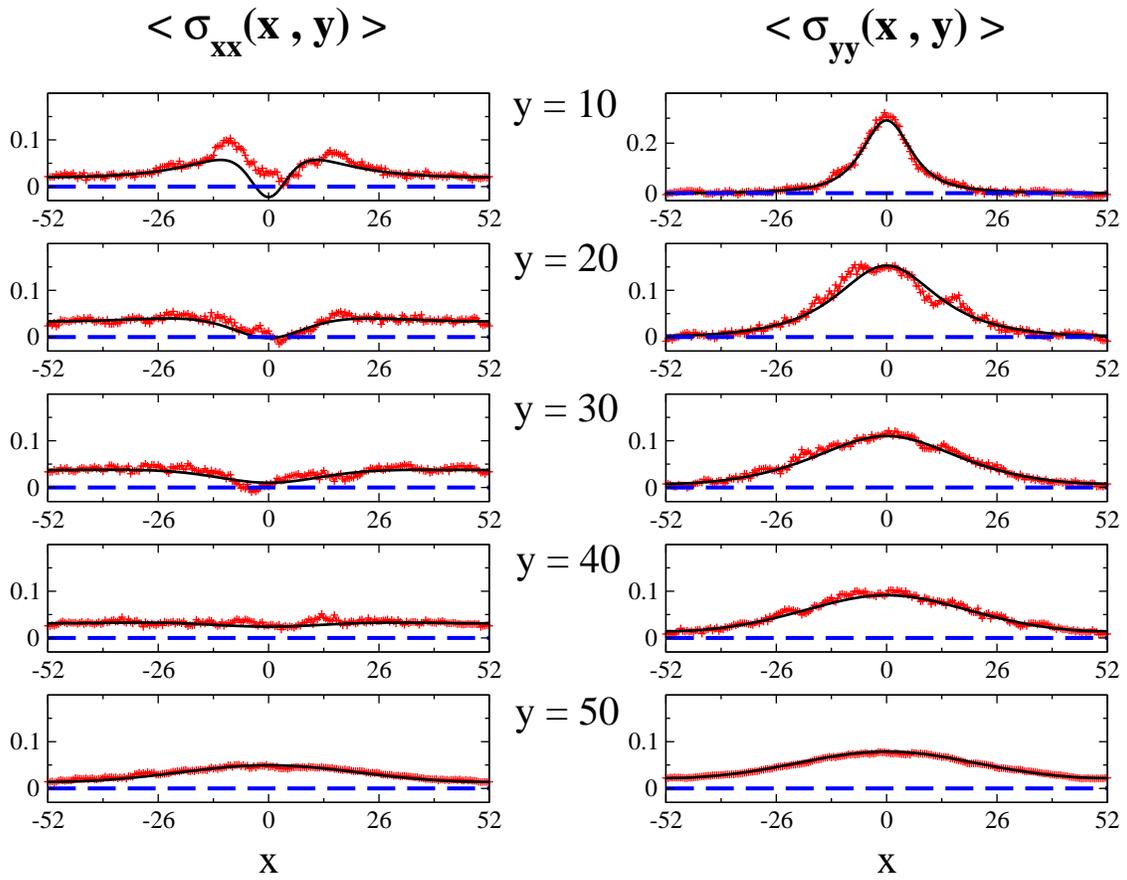}\end{center}
\vspace*{0.8cm}
\caption[]{(Color online)
Averaged incremental stresses \sigmaxx \ (left) and \sigmayy \ (right) versus $x$ for different vertical distances $y$ from the source. 
The boundary conditions indicated in Fig. 1(b) are used.
Data from configurations containing $N=10,000$ beads in boxes of $L=104$ is averaged over 220 independent measurements and compared with the predictions from classical elasticity (bold lines). 
The agreement is surprisingly good even for small $y$ and improves systematically with increasing source distance.
\label{figstressmean}}
\end{figure}

Two additional points have to be made here.
First, the responses compare already quite well with the analytical prediction (bold line) albeit they are not averaged over different realizations and despite the fact that the (not given) snapshot of the forces still looks quite noisy. This is obviously to be expected for large distances from the source as the response in an elastic body should self-average over the noise. While we shall make this more quantitative in a moment, Fig.~\ref{figstressD} shows clearly that a distance of order $y \approx 50$ yields a reasonable --- although not perfect --- self-averaged response.
This confirms our finding in \cite{papAmorph1} that systems of size $L=104$ show accurately the lowest eigenmodes and can therefore be regarded as free of finite size effects. This motivated our choice of this system size.
Note that the continuous response and the response averaged over many configurations (bold line in the Fig.~\ref{figstressD}) coincide at this distance from the point source (see also Fig.~\ref{figstressmean} on this point).

Second, we note in Fig.~\ref{figstressD} that the responses are identical for all systems where the forces per bead remain of order one or lower. This appears to be independent of the disk diameters $D$ despite the additional beads charged for larger disks. 
Apparently, these differences at the source are screened at $y=50 \gg D$. 
The response for $D=1$ and  $f_s/n_0 \approx 10$ is different, as the force per bead is outside the elastic regime. If we reduce the force per bead for $D=1$ further (filled circles) we obtain finally similar responses as for the larger disks. Note however, that linear response requires smaller forces per bead for smaller disks than for larger ones.

We now consider the mean stresses, i.e. the stress profiles averaged over many realisations (different samples and different application points of the force).
Far from the source, the self averaging discussed above implies that these mean profiles should behave in accordance with CET. This is less obvious close to the source, where fluctuations from one realisation to the other are large. In Fig.~\ref{figstressmean} we present the normal mean stresses $\la \sigmaxx \ra$ and $\la \sigmayy \ra$ as functions of $x$ for different vertical distances $y$. Similar curves have been obtained for the shear stress $\la \sigmaxy \ra$.
The agreement with CET (bold lines) is surprisingly good even for small distances from the source. It improves further with increasing distance $y$.
Apparently, the noise entering in the stress calculation is of (essentially) vanishing mean. While the vertical normal stress must have always one peak centered below the source, the horizontal normal stress is predicted by classical isotropic theory to show a minimum at $x=0$ between two peaks for $D \ll |y| \ll h$. This is a direct consequence of elasticity.
The double peak disappears close the fixed surface as there horizontal displacements which cause the tensile horizontal forces are suppressed.

\begin{figure}[t]
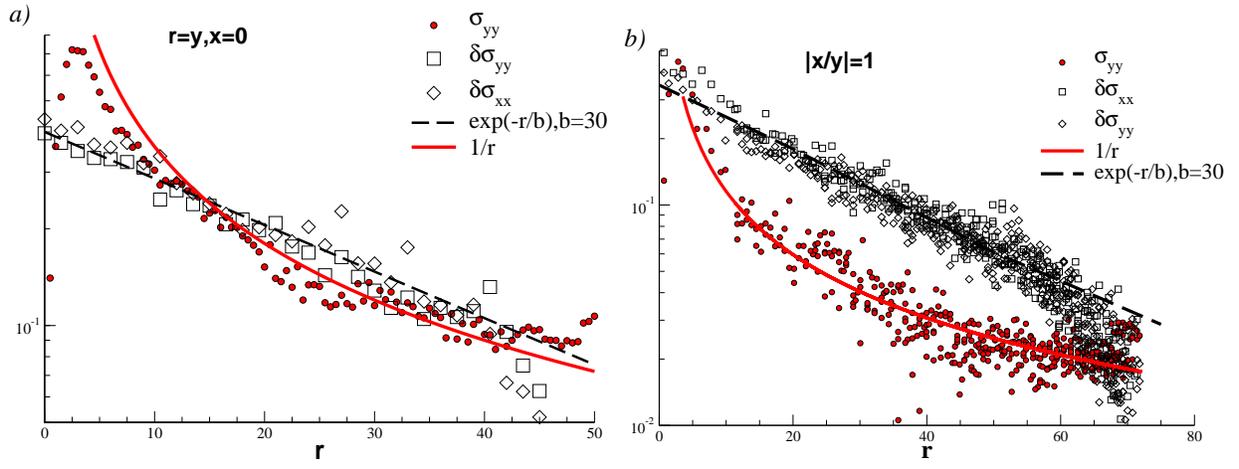

\begin{center}
\includegraphics*[width=0.49\textwidth]{{figselfaverage7a.eps}}
\includegraphics*[width=0.49\textwidth]{{figselfaverage7b.eps}}
\end{center}
\vspace*{0.8cm}
\caption[]{(Color online)
Comparison of the (incremental) stress fluctuations 
$\delta \sigma_{\alpha\beta} \equiv \left( \la \sigma_{\alpha\beta}^2\ra - 
\la  \sigma_{\alpha\beta} \ra^2 \right)^{1/2}$ 
with the mean vertical normal stress $\la \sigma_{yy} \ra$. 
This is done in two directions through the source:
(a) along the vertical line $(x=0)$ and (b) for $|x/y|=1$. 
We note that mean stresses and their fluctuations scale quite differently with distance $r$ from the source in both directions. For small distances we find relative fluctuations 
$\delta \sigma_{\alpha\beta}/\la \sigma_{\alpha\beta} \ra$ of order one. 
While the mean stresses decrease, as expected in 2D, essentially as $1/r$, the fluctuations are found to be well fitted by an exponential decay $\exp(-r/b)$ with $b \approx 30$. 
\label{figselfaverage}}
\end{figure}

As can be seen from Fig.~\ref{figselfaverage} for the normal stresses, all measured stresses decrease essentially as the inverse distance from the source (taken aside the expected corrections due the finite value of the system size). The two panels given in this figure correspond to measurements along two straight lines through the source: 
(a) $x=0$, (b) $x/y\equiv\tan(\theta)=\pm 1$.
Both figures look qualitatively similar.

More importantly, we compare in both panels both normal average stresses with their respective fluctuations from sample to sample
$\delta \sigma_{\alpha\beta} = 
\left( \la \sigma_{\alpha\beta}^2 \ra - 
\la \sigma_{\alpha\beta}^2 \ra^2 \right)^{1/2}$.
We note first that $\delta \sigma_{xx} \approx \delta \sigma_{yy} \approx \delta \sigma_{xy}$ (the latter relation not being represented in the figure) and that the fluctuations do not depend on the angle $\theta$ of the straight line, but solely on their distance $r$ from the source.
Surprisingly, we find fluctuations of order of the mean (normal) stresses, i.e. the relative fluctuations are of order one close to the source.
\footnote{For symmetry reasons, $\la \sigma_{xy} \ra = 0$ for $\theta \rightarrow 0$ and the corresponding relative fluctuation diverges.} 
This striking observation is by no means in conflict with the observed self-averaging far from the source, due to the different distance dependence of mean stresses and fluctuations. While the former decrease (essentially) analytically, our data suggests an exponential decay for the latter. Our fits are compatible with a characteristic screening length scale $b$ of order 30. 
Interestingly, this is of same order as the characteristic wave length $\xi$ we have found in \cite{papAmorph1} for the breakdown of the classical eigenmodes.
Only for distances somewhat larger than $b$, the self-averaging dominates over the analytical decay of the average stresses, and the {\em relative} fluctuations vanish eventually. 

In Fig.~\ref{fighistoforce}, we discuss in more detail the distribution of stresses along the $x=0$ line through the source. 
Only the vertical normal stresses $\sigma_{yy}$ are presented here, as the histograms for \sigmaxx \ and \sigmaxy \ show similar behavior.
The normalized histograms have been rescaled and plotted {\em versus} the natural scaling variable $u = \sigmayy /\la \sigmayy \ra$ which takes out the trivial distance dependence of the mean stress. Incidentally, as we know from Fig.~\ref{figstressmean}, we may equally use the analytically obtained stress as reference in the scaling variable, without changing the reduced histograms.

Three remarks have to be made here:
First, we note that all histograms scale reasonably well and the fluctuations, i.e. the width of the unscaled peaks, scale broadly as the mean stresses. Closer inspection reveals, however, that the rescaled peak width becomes slightly narrower with distance to the source. Both observations are obviously in perfect agreement with the previous Fig.~\ref{figselfaverage} where more or less constant relative fluctuation have been found due to the large value of the self-averaging length $b \approx h$. This masks somewhat the different functional dependency (analytic {\em versus} exponential) of the first two moments of the stress distributions.
Second, the distributions are more or less symmetric and the mean stress corresponds to the maximum of the histogram. This confirms the statement made above (Fig.~\ref{figstressmean}), that the fluctuations around the analytical prediction appear to be of vanishing mean. 
Third, although our statistics is certainly insufficient to characterize much better the shape of the distributions, specifically the scaling of their tails, a Gaussian distribution can be ruled out with the present data. In fact, as shown in the figure, an exponential fit is not unreasonable. In this sense the noise is large.

\section{Results for systems with compensation forces.}
\label{sec:resBY2}

While the previous section was mainly concerned about forces and stresses and their distribution, we will now, for the rest of this paper, investigate the displacement \usvec \ of the center of mass of the source region. This is the direct route to characterize the local elastic properties of an amorphous body. 
We use here open periodic boundary conditions without fixed particles, but with additional small compensation forces on all beads. As before, a vertically downwards pointing force acts on source disks of diameter $D=4$. 

\begin{figure}[t]
\rsfig{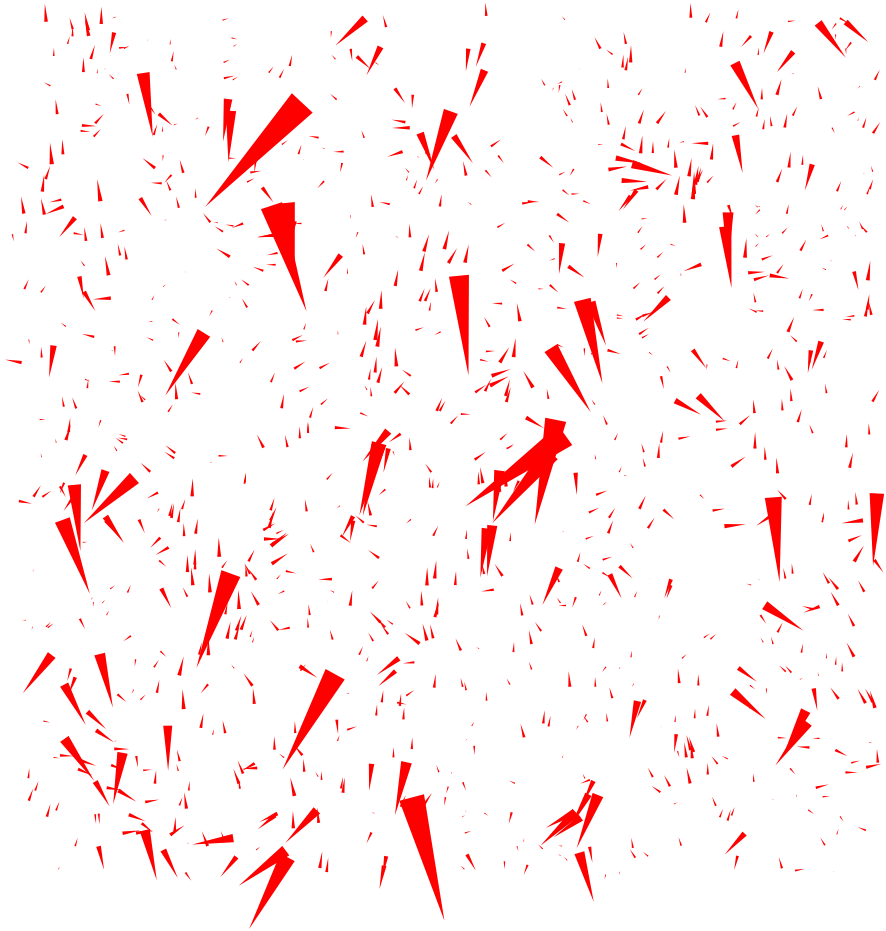}
\vspace*{0.8cm}
\caption[]{(Color online)
Snapshot of the reduced displacement vectors 
$\delta \usvec = \usvec - \la \usvec \ra$
of the center of mass of the source region. (The size of the arrows is proportional to the length of the reduced displacement vector.) This data have been obtained for completely periodic boundary conditions where no particles have been fixed. The vector field varies greatly in size and direction. Closer inspection shows strong spatial correlations.
\label{figsnapsourcedeltau}}
\end{figure}

Fig.~\ref{figsnapsourcedeltau} presents a typical snapshot of the noisy part $\delta \usvec = \usvec - \la \usvec \ra$ of the source displacements measured in one configuration.
For the given box size, the mean displacement substracted is roughly four times larger than the average fluctuation $\la \delta \usvec^{2} \ra^{1/2}$ (see Fig.~\ref{figcollectivity} below). Hence, the local elastic properties vary weakly with position. 
The snapshot (or a more detailed histogram) shows the bimodality of the $\delta \usvec$ distribution: Few very strong displacements point downwards in the direction of the force. They are due to some very soft spots.
The remaining $\delta \usvec$ are much smaller and strongly correlated in space. While pointing pretty much in all directions, they compensate obviously the net downward component of the soft spots.

We have checked the spatial correlations of the source displacements, by means of the (normalized) correlation function $ \la \delta \usvec(r) \cdot \delta \usvec(0) \ra$ which is summed over {\em all} pairs of displacements of a given configuration, and averaged over the configuration ensemble. In total, nearly 4000 responses contribute to the average correlation function presented in Fig.~\ref{figdisplcorr}. For very small distances, the correlation function should become constant since two sources of finite disk diameter excite the same beads. Equally, it is expected to become flat around $r\approx L/2$ due to the periodic boundary conditions. Both limits are in agreement with our data.
More importantly, the intermediate distance regime $D/2 \ll r \ll L/2$ may be reasonably fitted in log-log coordinates by a power law slope. Although a somewhat weaker value would even fit a larger range of the data we have indicated an exponent $-1$. Interestingly, the same dependency has been observed for the correlation function of the non-affine part of the displacement field discussed in Ref.~\cite{papAmorph1}. 
Finally, we remark that an additional non-analytic and possibly exponential regime for $b \ll r \ll L/2$ is conceivable for even larger boxes. This is suggested by the existence of the characteristic length $b$ observed in the self-averaging of the stresses. Unfortunately, as indicated by the broken line, this is at present not supported by our simulations due to the limited system sizes available. 

The reader will have observed that we do not probe here the rigidity of an isolated patch of material in a fixed frame. The fluctuations of the source displacements do not depend only on the local elements of the dynamical matrix but on a much larger neighborhood, in principle the whole system, whose effective size is estimated in the next section.

\section{Results for fixed topological layers.}
\label{sec:resBY3}

For solving the linear response directly by means of the dynamical matrix, it is useful to renumber and regroup the beads in topological layers around the source disk.
All the $n_1$ beads, interacting directly with the $n_0$ disk beads, are contained in the first neighbor layer, the $n_2$ beads interacting directly with the $n_1$ beads of the first neighbor shell are contained in the second layer, and so on (there are no direct interactions between the $n_0$ beads of the source and those of the second layer). Both the number of beads of each layer and the mean radius $R$ of the fixed particle layer around the source increase linearly with the topological rank from the source. 
The width of a topological layer is of order $3$ due to the cut-off of our potential and to the weak polydispersity \cite{papAmorph1}. The last layer of free particles containing $n_l$ beads, we fix the positions of the $N-(n_0+n_1+ ... + n_l) > 0$ remaining beads. 

With this renumbering, the structure of the dynamical matrix becomes more transparent, picturing systematically the influence of the subsequent topological layers.  
While in the last section {\em all} beads have been allowed to respond to the external load, we study here the effect of additional degrees of freedom when more and more topological layers and degrees of freedom are allowed to relax. The method used here is a systematic inversion of the dynamical matrix.

As in the last section, we only consider the displacement field at the source. In contrast, we use a source containing only one bead ($n_0=1$). The applied force is arbitrarily set to one.
Obviously, the response of the source, when all other beads are fixed ($l=0$), is directly described by the inverse diagonal coefficients of the dynamical matrix, which is a function of the local quenched forces and spring constants between neighboring monomers. 
For $l > 0$, the displacement $\UNvec$ of the source can be computed recursively by writing the equilibrium equations on all free beads. For $l=2$, for example, we get after combining all the $2\times (n_0+n_1+n_2)$ equilibrium equations in the two directions $x$ and $y$
\begin{equation}
\FNvec = \left( \Mtensor_{n_0 \times n_0} - \Mtensor_{n_0 \times n_1} \cdot \left(\Mtensor_{n_1 \times n_1} - \Mtensor_{n_1 \times n_2} \cdot \Mtensor^{-1}_{n_2 \times n_2} \cdot \Mtensor_{n_2 \times n_1} \right)^{-1} \cdot \Mtensor_{n_1 \times n_0} \right) \cdot \UNvec,
\end{equation} 
where $\Mtensor_{n_i \times n_j}$ is the matrix of size $2 n_i \times 2 n_j$ containing all the coefficients of the dynamical matrix relating the particles of the layer $i$ with the particles of the layer $j$. The ratio $u_s / f_s$, of the vertical component of the source displacement to the applied vertical force, can thus be easily computed, with the direct use of the coefficients of the dynamical matrix.

The vertical component of the source displacement $u_s$ is shown in Fig.~\ref{figcollectivity} as a function of the mean diameter $2R$ of the spherical region around the source which is allowed to respond to an external force. 
The open symbols correspond to the (reduced) mean displacement $\la u_s \ra/f_s$, the filled symbols to the fluctuation $\la \delta u_s^2 \ra^{1/2}/f_s$.
The squares are for the responses measured numerically after relaxation in periodic systems of linear size $L=2R=104$ without fixed beads (Sec.~\ref{sec:resBY2}). 
Note that the response at the source depends of course on $D$ for all $2R$.
\footnote{As different $D$ have been used for the two different boundary conditions, the results from the open periodic boundary method have been shifted vertically, taking into account the logarithmic correction suggested by continuum theory.}

We show, that the mean displacement increases logarithmically with system size $\la u_s \ra/f_s \approx 0.011 \log(2R/D)$ in qualitative agreement with continuum theory and this already for systems with only one free topological layer around the source ($l=1$). Hence, although $l=0$ is not sufficient, one can obtain the average local elastic moduli from a surprisingly small neighborhood region. Interestingly, the fluctuations level off at much larger distances of order of $b=30$ (corresponding to $l=6$ topological layers), as can be better seen from the inset. (It can be shown that the approach of the large system size limit is exponential.) This shows that, at system sizes of the order of the self-averaging length, the fluctuations become system size independent. For larger systems, $b$ determines the size of the region responsible for the noise in the source displacement field. 

\section{Concluding remarks.}
\label{sec:conclusion}

We have probed the incremental stress and displacement fields due to a point source force acting on two-dimensional amorphous Lennard-Jones solids. Focusing on the linear elastic response, this has been done for three different boundary conditions by means of the linearized Euler-Lagrange forces (i.e., the dynamical matrix) or by direct minimization of the total Hamiltonian. 

We demonstrate that the {\em average} stresses and displacement fields compare well with the predictions from classical isotropic elasticity, and this already for small distances from the source and for small system sizes (Figs.~\ref{figstressmean}, \ref{figcollectivity}). Contrasting to this, large stress (and, hence, strain) {\em fluctuations} are found for small distances to the source decreasing (essentially) exponentially with distance (Fig.~\ref{figselfaverage}). A surprisingly large length scale $b \approx 30$ is associated with this self-averaging with distance. Similarly, the fluctuations of the source displacement fields are found to become system size independent for $L=2R \gg 30$.
The self-averaging length $b$ is of the same order as the characteristic length scale $\xi$ associated with the non-affine displacement field under macroscopic strain setting the lower bound for allowing classical eigenfrequency calculation to be applicable. We believe that both length scales express the same physical fact and are indeed (up to prefactors) identical quantities.
This explains why vibration modes corresponding to wavelengths larger than $b$ (and, hence, averaging over larger distances) follow continuum theory, while modes with smaller wavelengths do not \cite{papAmorph1}.

As it has been pointed out, better statistics and much larger box sizes $L/2 \gg b \approx \xi$ would be required to establish unambiguously the scaling of the various distribution functions discussed here. Especially, an improved correlation function of the source displacements would be highly interesting to test the range of the observed power law (Fig.~\ref{figdisplcorr}). We strongly expect the existence of final cut-off at a characteristic length of order $b$. Additional theoretical guidance is also required to explain how such a large length scale arises for the fluctuations given the short-range correlation of the dynamical matrix elements \cite{papAmorph1}.
These results should be compared with experiments of nanoscale indentation on glasses~\cite{Loubet}, allowing a first experimental evidence of large scale fluctuations of the elastic properties in amorphous materials. The study of the pressure dependence of our results is currently in progress.

Coming back to the static properties of granular aggregates (composed of hard, cohesionless grains in frictional contact) invoked in the introduction, this work suggests several computer experiments.
As a first step, one should compare the average incremental stress and displacement responses. It may be interesting to verify if the double peak structure found for the horizontal normal stress is also present in granular matter although there the {\em total} normal stresses may not become tensile. (A priori this is possible since the {\em incremental} stresses are perfectly entitled to become negative as long as Coulomb's criterion is not violated.)
More importantly, the self-averaging properties of granular systems should be put to a test. Various experimental facts (especially for forces in vertical columns and silos) \cite{chaos} suggest much larger fluctuations with much weaker self-averaging properties compared to amorphous elastic bodies. 
Finally, it is a matter of debate, if the response to an arbitrary weak source does correspond to a Green's function in a strict mathematical sense. 
Additivity, linearity and reversibility of the responses should be tested directly. As the force network is subject to incessant restructuring due to the missing permanent grain contacts it may not be possible to describe --- even in the hydrodynamic limit --- the {\em total} charging of the packing as a {\em linear} operation \cite{chaos}.

\begin{acknowledgments}
During the course of this work, we had valuable discussions
from from P.~Claudin, E.~Duval, C.~Gay and E.~Kolb. 
Computational support by IDRIS/France and CEA/France is acknowledged.
\end{acknowledgments}

\appendix
\section{Two-dimensional elastic response to a point force.} 
\label{sec:appendix}

At equilibrium the stress state of a two-dimensional elastic material must satisfy the (two) force balance equations
$\nabla_i\sigma_{ij}=F\delta (x)\delta (y)$
with $i,j=x,y$ and  $F$ being the external point force applied in $(0,0)$.
The main assumption of classical elasticity \cite{Salencon} is that the system may be entirely described by the continuous displacement field $\vec{u}(x,y)$. Due to global translational and rotational invariance, only the strain field --- by definition the symmetric part of the gradient of the displacement field --- appears in the equations. Moreover in {\em linear} elasticity, the stress tensor is related to the strain tensor $\epsilon_{ij}$ through Hooke's law.
Only two phenomenological parameters are required for isotropic homogeneous systems, $E$ and $\nu$ (or $\lambda$ and $\mu$). Supposing this, it follows the compatibility equation \cite{Salencon}  
$\partial_{xx}\sigma_{yy}-\nu\partial_{yy}\sigma_{yy} + \partial_{yy}\sigma_{xx}-\nu\partial_{xx}\sigma_{xx}=2(1+\nu)\partial_{xy}\sigma_{xy}$
and thus, combined with the force balance equations, the well-known Laplace equation 
$\triangle (\sigma_{xx}+\sigma_{yy})=0$ for $(x,y)\neq (0,0)$.

\begin{figure}[t]
\rsfig{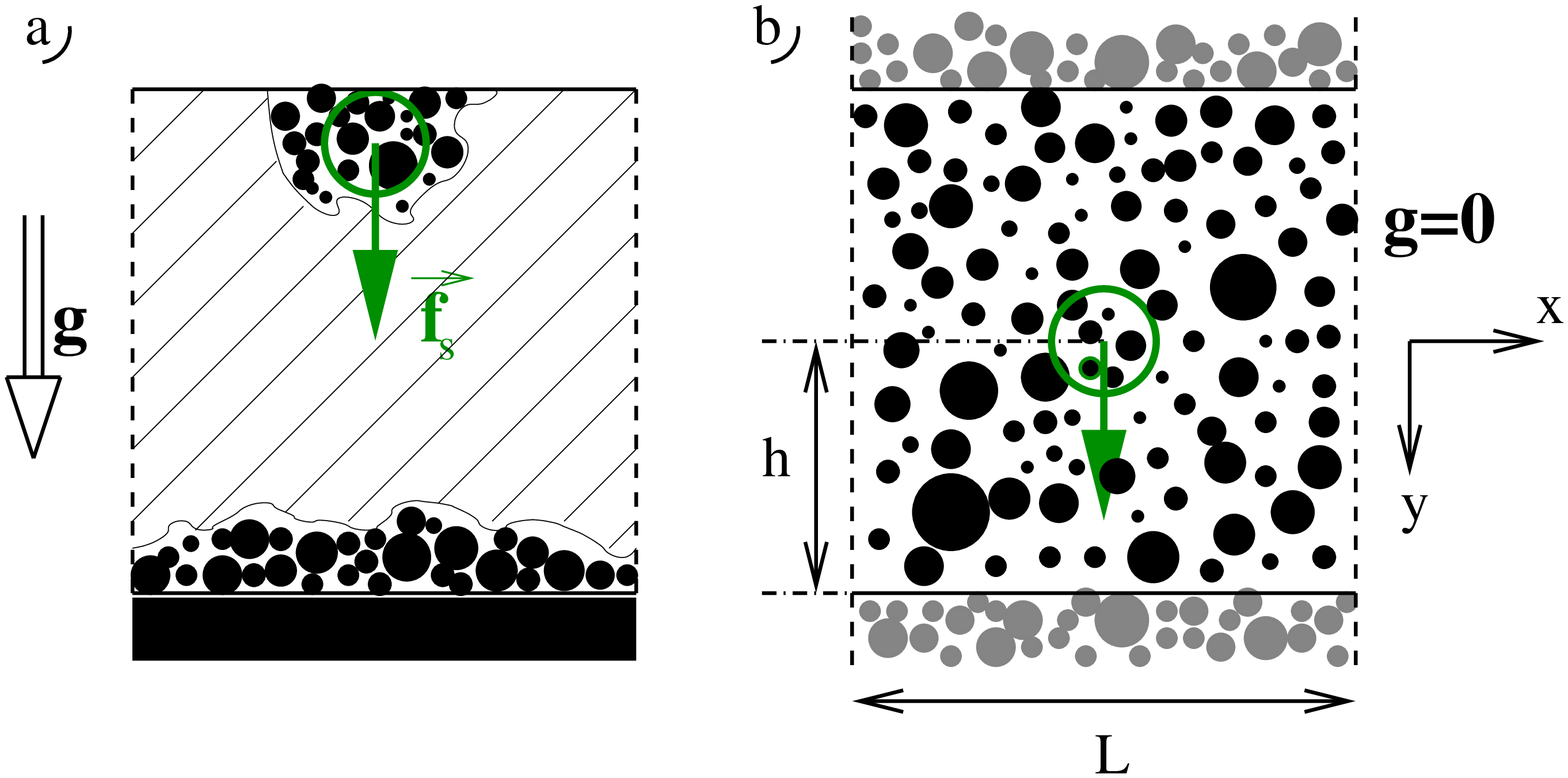}
\vspace*{0.8cm}
\caption[]{ (Color online)
Sketch of two boundary conditions of interest for measuring the response to an additional point force source \fsvec :
(a) The source may be applied to the free upper surface of a prestressed aggregate formed at constant gravity on a rigid bottom plate (possibly containing some stress transducers). This setup has been studied extensively recently \cite{chaos,claudinaniso,clement} in order to determine the static response of packings of (hard and cohesionless) granular matter.  
(b) One of the three boundary conditions studied in this paper. The source is applied within a {\em macroscopically} isotropic and homogeneous ``computer solid" in a periodic simulation box of linear size $L$. The center of the source defines the origin of the coordinate system $(x,y)$. For mechanical stability we either apply a compensation force of $-\fsvec/N$ ($N$ being the total number of beads) to all particles or we freeze some particles (gray beads) as shown in the right panel. 
We study the response of amorphous packings of carefully quenched (slightly polydisperse) Lennard-Jones beads. Obviously, this is a further simplification with regard to the granular material case with its more intricate non-linear (static friction) particle interactions.
\label{figsketch}}
\end{figure}

As a specific example we present here the calculation for the point source problem between two fixed horizontal walls posed in Fig.~\ref{figsketch}(b). Hence, the displacement field $\vec{u}$ must vanish for $|y|=h$. Periodicity and symmetry of the simulation box in horizontal direction impose a solution periodic and symmetric (odd or even) in $x$. These boundary and symmetry conditions can be readily reformulated in terms of the stresses. 

To obtain the elastic response, the idea is to use the method presented in ~\cite{claudinelastic} in the case of an elastic layer submitted to a force localized at its surface. We divide our medium into two parts: part 1 above, part 2 below the point source. The continuity of the displacement field along the fictitious dividing line requires
$u_x^{(1)}(x,0)=u_x^{(2)}(x,0)$ and
$u_y^{(1)}(x,0)=u_y^{(2)}(x,0)$.
The point force is taken into account by imposing
$\sigma_{yy}^{(1)}(x,0)=\sigma_{yy}^{(2)}(x,0)-p(x)$
where $p(x)$ is the vertical external pressure.
An additional constraint is imposed by the continuity of the shear stress 
$\sigma_{xy}^{(1)}(x,0)=\sigma_{xy}^{(2)}(x,0)$.
Note that the continuity of $u_x$ at $y=0$ together with the discontinuity of $\sigma_{yy}$ imposes the discontinuity of $\sigma_{xx}$. Imposing a continuous $\sigma_{xx}$ would yield to a discontinuous displacement field $u_x$ with large scale vortices~\cite{papAmorph4}.

The stress tensor components are decomposed into a base of harmonic functions, typically affine functions or product of trigonometric functions with exponentials. Taking into account the $x \longleftrightarrow -x$ symmetry, we look for a solution of the type
\begin{eqnarray}
\sigma_{xx}^{(1,2)}+\sigma_{yy}^{(1,2)} &=& \sum_{n=0}^{+\infty} \cos(qx) 
\left(a^{(1,2)} e^{qy} + b^{(1,2)} e^{-qy}\right) \nonumber\\
\sigma_{xx}^{(1,2)}-\sigma_{yy}^{(1,2)} &=& 
\sum_{n=0}^{+\infty} \cos(qx) \left(qy \left(a^{(1,2)} e^{qy} - b^{(1,2)} e^{-qy}\right) + 2\left(c^{(1,2)} e^{qy}-d^{(1,2)} e^{-qy}\right)\right)\nonumber\\
\sigma_{xy}^{(1,2)} & = & \sum_{n=0}^{+\infty} \sin(qx) \left( qy/2 \left(a^{(1,2)} e^{qy} + b^{(1,2)} e^{-qy}\right) + \left(c^{(1,2)} e^{qy} + d^{(1,2)} e^{-qy}\right)\right)\label{eq:sig}. 
\end{eqnarray}
$a^{(1,2)}$,$b^{(1,2)}$,$c^{(1,2)}$ and $d^{(1,2)}$ are coefficient functions depending on the frequency $q = n \Delta q$. The latter is quantized with $\Delta q\equiv 2\pi/L$, due to the finite horizontal width $L$ of the layer, and periodic boundary conditions. A similar looking ansatz for the corresponding displacement fields is readily obtained.

The vertical external pressure is expressed in the same form, as
\begin{equation}
p(x)=\sum_{n=0}^{+\infty} \cos(qx)s(q).\Delta q
\nonumber
\end{equation} 
with $p(x)$ either equal to a gaussian $\left(s(q)=F/\pi e^{(-q^2 a^2/2)}\right)$, or a uniform force of width $2a$ $\left(s(q) = F \sin(qa)/(\pi a q)\right)$.

Replacing the general expressions for the stresses and the displacement fields in the equations characterizing the boundary conditions and the continuity at the fictitious dividing line, leads to the following explicit expressions for the eight functions $a^{(1,2)}$, $b^{(1,2)}$, $c^{(1,2)}$ and $d^{(1,2)}$ depending on $s(q)$, $qh$ and $\nu$
\begin{eqnarray}
a^1(q) &=& -\frac{s(q)\Delta q}{4}\frac{\left((1+\nu).2qh+3-\nu\right)e^{-2qh} + (\nu-3)e^{-4qh}}{D(q)}\nonumber\\
a^2(q) &=& \frac{s(q)\Delta q}{4}\frac{\left((1+\nu).2qh-3+\nu\right)e^{-2qh} - (\nu-3)}{D(q)}\nonumber\\
b^1(q) &=& \frac{s(q)\Delta q}{4}\frac{\left(-(1+\nu).2qh+3-\nu\right)e^{-2qh} + (\nu-3)}{D(q)}\nonumber\\
b^2(q) &=& \frac{s(q)\Delta q}{4}\frac{\left((1+\nu).2qh+3-\nu\right)e^{-2qh} + (\nu-3)e^{-4qh}}{D(q)}\nonumber\\
c^1(q) &=& \frac{s(q)\Delta q}{8(1+\nu)}\frac{\left(2(1+\nu)^2q^2h^2+(1-\nu^2)2qh+(1-\nu)(3-\nu)\right)e^{-2qh} + (1-\nu)(\nu-3)e^{-4qh}}{D(q)}\nonumber\\
&=& d^2(q)\nonumber\\
c^2(q) &=& \frac{s(q)\Delta q}{8(1+\nu)}\frac{\left(2(1+\nu)^2q^2h^2-(1-\nu^2)2qh+(1-\nu)(3-\nu)\right)e^{-2qh} + (1-\nu)(\nu-3)}{D(q)}\nonumber\\
&=& d^1(q)\nonumber\\
{\rm with}\, D(q) &\equiv & {2qhe^{-2qh}-\frac{\nu-3}{2(\nu+1)}+\frac{\nu-3}{2(\nu+1)}e^{-4qh}}\nonumber
\end{eqnarray}

Substituting these coefficient functions back into the general ansatz for stress and displacement fields, one obtains explicit expressions for the stress and the displacement fields. We have drawn numerically these expressions to get the theoretical fits presented in Sec.~\ref{sec:resBY1}. Note that unlike the boundary condition studied in \cite{claudinelastic}, the coefficient functions depend now on $\nu$. The displacement field $\vec{u}$ is proportional to  $1/E$. The solution also depends on the system height $2h$, on the size $a$ of the source, and on the width $L$, the latter due to the quantization of the Fourier integration.



\newpage

\newpage

\newpage

\newpage

\newpage

\newpage

\newpage
 
\newpage
\begin{figure}[t]
\rsfig{fighistostress8.eps}
\vspace*{0.8cm}
\caption[]{(Color online)
Normalized distributions of incremental vertical normal stresses for different distances $y > 0$, along the vertical line through the source ($x=0$).
The histograms are plotted versus $u = \sigmayy/\la \sigmayy \ra$. The data is averaged over 220 independent point source experiments. The histograms are more or less symmetric. With increasing distance, the rescaled distributions become systematically narrower, in agreement with the previous figure, however, the effect is weak. For $y < 40$, the tails of the histograms are not Gaussian, but roughly exponential as indicated by the bold lines.
\label{fighistoforce}}
\end{figure}

\newpage

\newpage
\begin{figure}[t]
\rsfig{figdisplcorr10.eps}
\vspace*{0.8cm}
\caption[]{(Color online)
Spatial correlation function  
$\la \delta \usvec (r) \cdot \delta \usvec (0) \ra/ \la \usvec^2 \ra$ of the source displacement vector $\delta \usvec = \usvec - \la \usvec \ra$ with $r$ being the distance between source terms within the same configuration. Note that the correlation function is normalized. The average is taken over a total number of nearly 4000 linear responses using open periodic boundary conditions without fixed particles. As indicated by the bold line, the correlation function decreases essentially like the inverse distance for $D/2 \ll r \ll L/2$. It becomes constant for smaller and larger distances. 
We strongly suspect an additional exponential cut-off (dashed line), however, our data is too noisy and, more importantly, $L$ is too small to demonstrate this unambiguously.
\label{figdisplcorr}}
\end{figure}

\newpage
\begin{figure}[t]
\rsfig{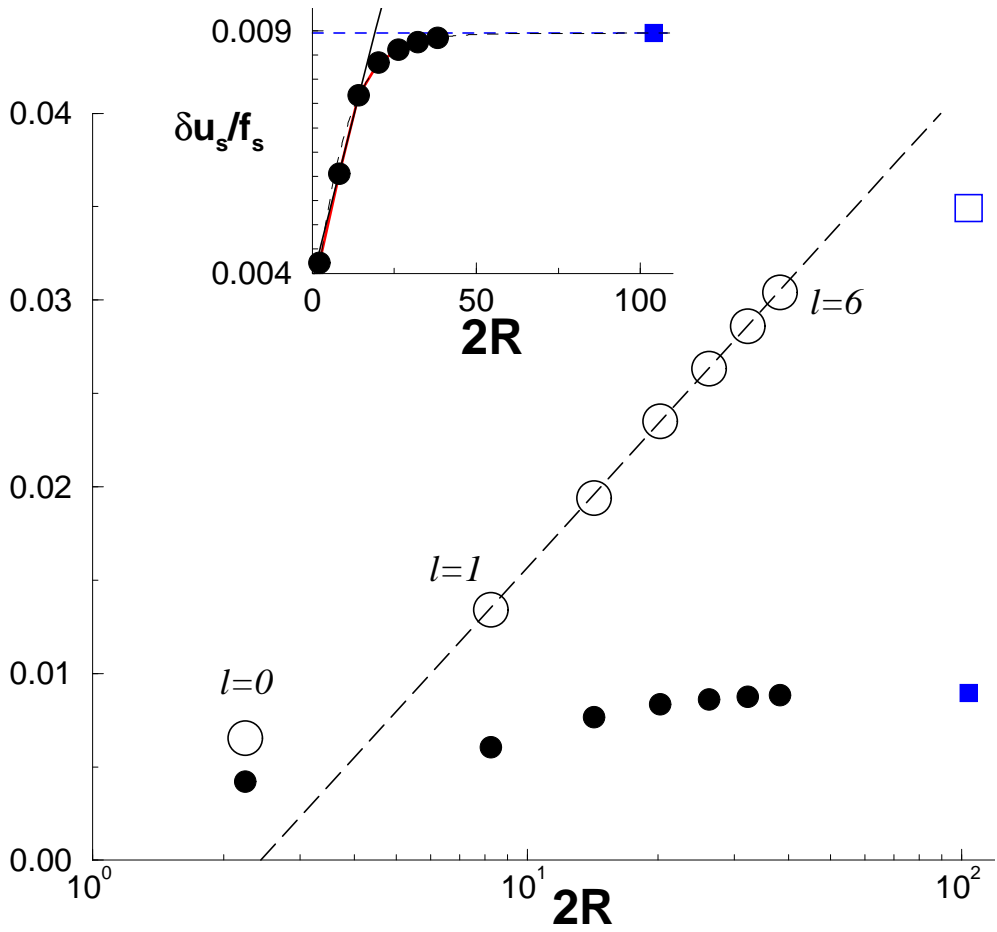}
\vspace*{0.8cm}  
\caption[]{(Color online)
Vertical component of the displacement of the source center of mass $u_s$ for a given total force $f_s$ as a function of system size $2R$. The open symbols correspond to the mean displacement $\la u_s ^2 \ra^{1/2}/f_s$, the filled symbols to the fluctuation $\la \delta u_s^2 \ra^{1/2}/f_s$. The circles are for the boundary condition with fixed topological layers discussed in Sec.~\ref{sec:resBY3}, $R$ being the mean distance to the fixed border shell. The squares are for responses in periodic systems of linear size $L=2R=104$ without fixed beads (Sec.~\ref{sec:resBY2}).
The mean displacement increase logarithmically with system size $\la u_s \ra/f_s \approx 0.011 \log(2R/D)$ in agreement with theory. The fluctuations level off at distances of order of $b=30$, as can be better seen from the inset. 
\label{figcollectivity}}
\end{figure}

\end{document}